\documentclass[reprint,nofootinbib,
 amsmath,amssymb,
pra,
]{revtex4-2}
\usepackage{url,hyperref,lineno,microtype,subcaption,listings,xcolor,graphicx,verbatim}
\usepackage[finalnew]{trackchanges}
\usepackage{booktabs}
\addeditor{DWH}

\definecolor{codegreen}{rgb}{0,0.6,0}
\definecolor{codegray}{rgb}{0.5,0.5,0.5}
\definecolor{codepurple}{rgb}{0.58,0,0.82}
\definecolor{backcolour}{rgb}{0.95,0.95,0.95}

\lstdefinestyle{mystyle}{
    backgroundcolor=\color{backcolour},   
    commentstyle=\color{codegreen},
    keywordstyle=\color{magenta},
    numberstyle=\tiny\color{codegray},
    stringstyle=\color{codepurple},
    basicstyle=\ttfamily,
    breakatwhitespace=false,         
    breaklines=true,                 
    captionpos=b,                    
    keepspaces=true,                 
    numbers=left,                    
    numbersep=5pt,                  
    showspaces=false,                
    showstringspaces=false,
    showtabs=false,                  
    tabsize=2
}

\begin{document}
\title[Recategorising research]{Recategorising research:\\Mapping from FoR 2008 to FoR 2020 in Dimensions}

\author{Simon J Porter}
 \email{s.porter@digital-science.com}
 \affiliation{%
 Digital Science, 6 Briset Street, London, EC1M 5NR, UK
}
\author{Lezan Hawizy}
 \email{l.hawizy@digital-science.com}
 \affiliation{%
 Digital Science, 6 Briset Street, London, EC1M 5NR, UK
}
\author{Daniel W Hook}%
\email{daniel@digital-science.com}
\affiliation{%
 Digital Science, 6 Briset Street, London, EC1M 5NR, UK
}%
\affiliation{Centre for Complexity Research, Imperial College London, London, SW7 2AZ, UK}
\affiliation{Department of Physics, Washington University in St Louis, St Louis, Missouri, US.} 

\begin{abstract}
In 2020 the Australia New Zealand Standard Research Classification Fields of Research Codes (ANZSRC FoR codes) were updated by their owners.  This has led the sector to need to update their systems of reference and has caused suppliers working in the research information sphere to need to update both systems and data.  This paper focuses on the approach developed by Digital Science's Dimensions team to the creation of an improved machine learning training set, and the mapping of that set from FoR 2008 codes to FoR 2020 codes so that Dimensions classification approach for the ANZSRC codes could be improved and updated.
\end{abstract}

\maketitle

\section{Introduction}
\label{sec:1}
In 2020 the organisations behind the Australian and New Zealand Standard Research Classification (ANZSRC)---the Australian Bureau of Statistics (ABS), Stats NZ, the Australian Research Council (ARC) and the New Zealand Ministry of Business, Innovation and Employment (MBIE), completed a review of all three components of the classification, including the Fields of Research (FoR) Codes. Over recent years the FoR codes have become an increasingly popular classification system due to their broad subject coverage and well-formulated three-level structure.  The FoR codes have been used extensively in Digital Science products such as Altmetric, Symplectic Elements, Figshare and Dimensions.  In particular, Dimensions makes use of machine-learning approaches to classify research objects across its database at a per object level rather than at a journal level.  This was a logical requirement of the Dimensions database as it includes not only publications but also grants, patents, policy documents, clinical trials and datasets, none of which come with a journal ``wrapper'' and which, for the purposes of analysis, must have a standardised classification approach applied across them \cite{hook_dimensions_2018}.  When the change in coding was announced it was clear that a mapping from the FoR 2008 Codes to the FoR 2020 Codes would be needed and this opened up the opportunity to revisit the machine-learning training set that had originally been used to develop the FoR 2008 classification technology.  This paper focuses on explaining the methodology for the creation of a new machine-learning training set for Dimensions.

It is perhaps unusual to focus on the methodology for the creation of a machine-learning training set in an academic paper.  However, the creation of training sets is something of a ``dark art'' and with the rising use of AI and machine learning approaches becoming increasingly characteristic in scholarly information infrastructure we argue that greater transparency is in the interests of the communities served by these systems. Calls for responsible research evaluation \cite{fochler_implicated_2017,gadd_influencing_2019, gadd_mis-measuring_2021,szomszor_interpreting_2021} need also to extend to transparency in these types of algorithm as they are ultimately behind the setting of reference frames for comparison groups and for benchmark averages, and ultimately support important decisions around funding and careers. 

We do not provide a detailed analysis of precision-versus recall of the results of our mapping, nor do we focus on the machine-learning component of this treatment. Rather, we feel that the the novelty in this presentation centres around the various strategies used to create a training set that allows the problem of mapping to be accessed in an environment of sparse data. Thus, this paper contains no detailed analysis concerning the validation of the Dimensions FOR categorisation system.  However, we do feel that the treatment presented here will be of interest to those who wish to create datasets to support machine learning and can, perhaps be an inspiration for approaches that can yield data in situations where it is hard to come by.

\subsection{The importance of classification systems}
\label{sec:1.1}
The assignment of subject areas to a piece of research is not simply an act of classification, but rather it is an important step in setting the work in its context. The selection and application of subject areas to label a research output provides a ``lens'' through which to understand the  relationships and embedding of ideas included in the work in broader scholarly contexts.  They also help to place the context and lineage of the authors and institutions involved in the work, and of the funders of the work.  Over the centuries, they have  provided a route to the creation of communities of practice and through that the establishment of standards and norms for a field, as well as a frame through which to compare, contrast and evaluate work.  Thinking of a categorisation scheme as a piece of the measurement equipment of research evaluation is, we argue, an apt analogy.

Looking through the classification lens in the opposite direction allows us to recognise that a classification or categorisation scheme also tells us implicitly about those undertaking the classification \cite{szomszor_interpreting_2021}. Is the scheme defined coarsely, at a high level, or is detailed and fine-grained?  Is it multi-layered?  Understanding  the criteria that define the scope and nature of the classification scheme allows us to infer at least some of the needs and uses of the body carrying out the classification.  Whether the scheme covers subject areas evenly or contains intentional (or unintentional) biases; whether it is subject-led or grand-challenge-led, can be telling as to the motivations or worldview of those defining the scheme.

The notion that classification schemes are passive and immutable definitions of subject areas is one that is easy to fall into.  After all, our early experiences of high-level classifications are things like ``Physics'', ``Chemistry'' and ``Biology'', which appear to us to be solid, well-differentiated areas.  However, differentiating between Physics and Chemistry in a research context can often be challenging (as evidenced for example by the crossover in Nobel Prize winners between the two subjects) and working out where Chemistry ends and Biology begins, and what constitutes Biochemistry, is no less challenging.  In addition, as research continues into these fields our store of knowledge grows, our definitions develop, new fields are founded, fields merge and older areas are subsumed or deprecated in favour of more modern understandings \cite{zitt_bibliometric_2019}. It may also be argued that, in today's world, more challenge-driven or outcome driven classification systems should be held more fundamental than the classic subject definitions.  To this end, there has been significant work in developing approaches to SDG classification such as {\cite{purnell_comparison_2022}}.

From an evaluative perspective the dynamical nature of the classification of knowledge and  scholarship can lead to challenges. Typically, in contexts where policies or strategies are being developed, initiatives being evaluated or reports being compiled for external stakeholders, the reference frame needs to be held constant to provide a comprehensible “like-for-like” comparison.  From this basis benchmarks can be constructed, trends observed, and comparisons made between different individuals, institutions, countries or subjects.  This requires stability of classification.

Given the central nature of classification structures in reporting to key stakeholders it is unsurprising not only that they are widely used, but also that they are highly embedded into the academic ecosystem.  From their use in national evaluation exercises, to their inclusion in research information management systems and internal university modelling exercises.  More subtly, the prevalent classification system implicitly defines how we think about subject funding and the very notion of interdisciplinarity.  Given that the classification system is part of the measurement equipment, it is clear that measurements of a system have tangible effects on the things that are being measured \cite{goodhart_problems_1981,strathern_improving_1997}.  

For infrastructure providers, who need work to provide homogeneous systems and data across subjects and geographies, this complex landscape creates a whole plethora of challenges.  There is a natural tension between homogeneity ( hence usability) and specificity ( hence the value to individual users) of systems and classifications schemas.

All this means that the choice of a classification system by a research funder often codifies their priorities and hence ingrains those priorities in the users of the system.  Consequently, there is a considerable amount of work in changing a system once chosen.\footnote{Dimensions was released in its current form in 2018 and the choice of the classifications that were implemented for that release were heavily influenced by the systems prevalent in other tools in the Digital Science family (in particular Symplectic Elements and Figshare) that had, in turn, been influenced in their choices by their teams' experience in the Australian sector.  Since Digital Science's products are used by internationally there are many institutions who make extensive use of ANZSRC FoR codes in their work that are beyond Australia.  Hence, from a Digital Science perspective, updating the ANZSRC FoR codes has significant impacts not just for the Australian sector.} That work is not limited merely to defining the classification but also to the systematic changes both infrastructural, cultural and psychological that result from that change.  At the same time there is an inbuilt tension between a classification that is up-to-date and hence reflective of and appropriate for the measurement of research, and the disruption and work associated with the implementation of a new system.

\subsection{The ANZSRC system}
\label{sec:1.2}
Australia and New Zealand have, for many years, worked together on the definition and implementation of a common research classification system known as the Australian and New Zealand Standard Research Classification (ANZSRC) Fields of Research Codes (or simply the FoR codes).   The FoR codes are one part of the ANZSRC system, which also includes classification schemas for Type of Activity (TOA) and Socio-economic Objectives (SEO).  The FoR Codes comprises a three-level hierarchical classification system: The highest level are labelled by 2-digit codes (i.e. a 2-digit number) where each code refers to a “Division” level classification, describing a whole high-level subject area such as ``Biomedical and Clinical Sciences''; the middle level is labelled by 4-digit codes (i.e. a 4-digit number) and characterise a ``Group'' level that describes groups of fields such as “Cardiovascular medicine and haematology”; the lowest level is labeled by 6-digit codes where a quite specific field is codified such as ``Haematology'' \cite{arc_australian_2020}.

The most recent collaboration started in before 2008 as a joint project between Australian Bureau of Statistics (ABS), Statistics New Zealand (Stats NZ), Australian Research Council (ARC) and the New Zealand Ministry of Business, Innovation \& Employment.  Since its original inception the project has released two versions of a classification system: the first in 2008 and the most recent being released in 2020, this was pre-dated by two releases of the Australian Standard Research Classification (ASRC) one in 1993 and another in 1998.  It was originally designed to classify Australian and New Zealand-based research outputs for the purposes of tracking national research activities and making strategic funding and policy decisions.

The scheme is based on the structure of the Frascati manual \cite{oecd_frascati_2015} but its granularity and subject biases are based on a survey of research output from Australia and New Zealand.  Thus, this particular piece of ``measurement machinery'' can be regarded as a custom-made tool designed specifically to measure Australia and New Zealand with respect to a given reference point.  One important output of this has been a journal list that specifies classification codes for each scholarly journal.  This ``codex'' has allowed research funding organisations in Australia and New Zealand to guide the classification of work included in evaluation processes.

When Dimensions was launched in 2018, it moved away from journal-level classification as it included different types of research objects including books, datasets, policy documents, grants, patents and clinical trials, none of which are contained in journal wrappers.  Thus, in order to produce a homogeneous experience for users, an object-level classification was put in place.  This was made possible by a combination of the use of machine-learning technology and access to full text, which allowed automated classification of objects in the database.  There are significant advantages to this approach since, with a given, static, learning data, one can ensure that classification is consistent across a long time period (something that is impossible to do with human classifiers since they cannot help but learn - this gives rise to a natural drift in classification that then makes it difficult to argue that one is reporting on a like-for-like basis to stakeholders).  But, it also comes with a major challenge in that one needs a well-annotated dataset in order to provide a reference dataset for the machine learning algorithms.

\subsection{Dimensions implementation}
\label{sec:1.3}
The Digital Science Dimensions team implemented several different classification systems using machine-learning approaches as part of the initial release of Dimensions in 2018 \cite{hook_dimensions_2018,herzog_dimensions_2020, wastl_expanding_2020}.  Many of these centred around health and disease classifications since there are many analytics and evaluative use cases that call on these classifications.  The ANZSRC FoR Codes 2008 were implemented as the one ``universal'' classification across all topics.  Due to the body of training data available, it was not possible to classify research objects in Dimensions reliably at a 6-digit level and hence the 4-digit level was implemented, with 2-digit classifications derived from the 4-digit level classifications assigned to an object.  In this process, each object in Dimensions is capable of receiving multiple classifications.

The current paper focuses on the process that Digital Science has taken to building a replacement for the dataset and machine learning approach that we used to train the classification of the 2008 FOR codes, with the approach for the new 2020 FOR codes.  The paper is arranged as follows: In section~\ref{sec:2} we describe our methodology; in section~\ref{sec:3} we give an overview of the new mapping; and in section~\ref{sec:4} we discuss possible next steps and further approaches to improve our mapping.

\section{Methodology}
\label{sec:2}
When creating a classification system afresh, there are typically fewer degrees of freedom than one might expect.  One must choose the level of granularity and what kind of representation of meso-level structures might be appropriate for the intended use cases.  Yet, there is usually significant freedom in methodology: different clustering approaches provide different outcomes \cite{small_structure_1974,wagner_new_2007, subelj_clustering_2016}.  The problem here is more constrained - we need to create a training set that maps to a specific target structure that has already been defined.  In addition, we need to map that target structure to a new target structure and ensure consistency.  The resultant classification and the mapping needs to make sense and provide a narrative that makes sense to users of the classification. 

Helpfully, when the new version of the ANZSRC FoR codes were released in 2020, a mapping was provided so that 80\% of the 2008 codes could be mapped directly to codes in the new 2020 classification.  The objects in Dimensions already associated with these codes could be moved directly to the new codes.  However, for the 20\% of codes that were either new or were splits or conglomerations of existing codes a new approach needed to be found.  Naturally, as the new codes were exactly that, new, there was no reference dataset with items that had been annotated and labelled.

The process that we detail below was taken across 100\% of all codes - the 80\% of the existing ``lift and shift’’ codes were used as a reference/test dataset.  So, all outputs in Dimensions have been mapped using the new methodology described below.  This was done in order to ensure homogeneous treatment across fields.

We start the process by taking a reference dataset of ARC grants that have been annotated with 2008 FoR Codes, which has been available for some years \cite{arc_grants_2022,nhmrc_outcomes_2022}.  This dataset was the original dataset that we used as the basis for the learning models for the original FoR coding in Dimensions.  This dataset has four core drawbacks:

\begin{enumerate}
    \item The dataset is associated with grants, which poses several problems including the small size of texts available for analysis and the more limited language usage that tends to be characteristic of word-limited grant proposals;
    \item The dataset is small - across all years available from both the ARC and NHMRC, the total size of the dataset is just over 60,000 records.
    \item There is an implicit bias in these data toward how research is framed through the lens of the priorities of Australia’s funding priorities. While, for the purposes of Australia and New Zealand evaluations and category handling, this may have been a natural or even desirable bias.  However, the ANZSRC system has gained a lot of support internationally, being used in several products (such as Dimensions).  Additionally, the Australian and New Zealand research systems are very much a part of the global research community, and hence it makes sense that a categorisation system that is so broadly used should have a solid international perspective.  We note that the dataset is annotated at the 6-digit level for these data - which gives greater granularity than needed for our purposes here, but which is nonetheless helpful in checking our approach.
    \item Finally, at the time of developing this methodology the grant dataset is annotated with 2008 codes and not 2020 codes, and ultimately we need to have a training set that is annotated with 2020 codes.
\end{enumerate}

We employ four strategies to address the limitations of the existing dataset.  The first three strategies address the first three points raised above, resulting in a larger, less geographically biased, publication-based training set annotated with FoR 2008 codes at a 4-digit level.  The final strategy maps these annotations to the FoR 2020 codes at a 4-digit level.  This resultant dataset can then be used as a training dataset in the Dimensions machine learning algorithm.

\subsection{Strategy 1: From Grants to Publications}
\begin{figure*}
    \centering
    \includegraphics[width=.9\linewidth]{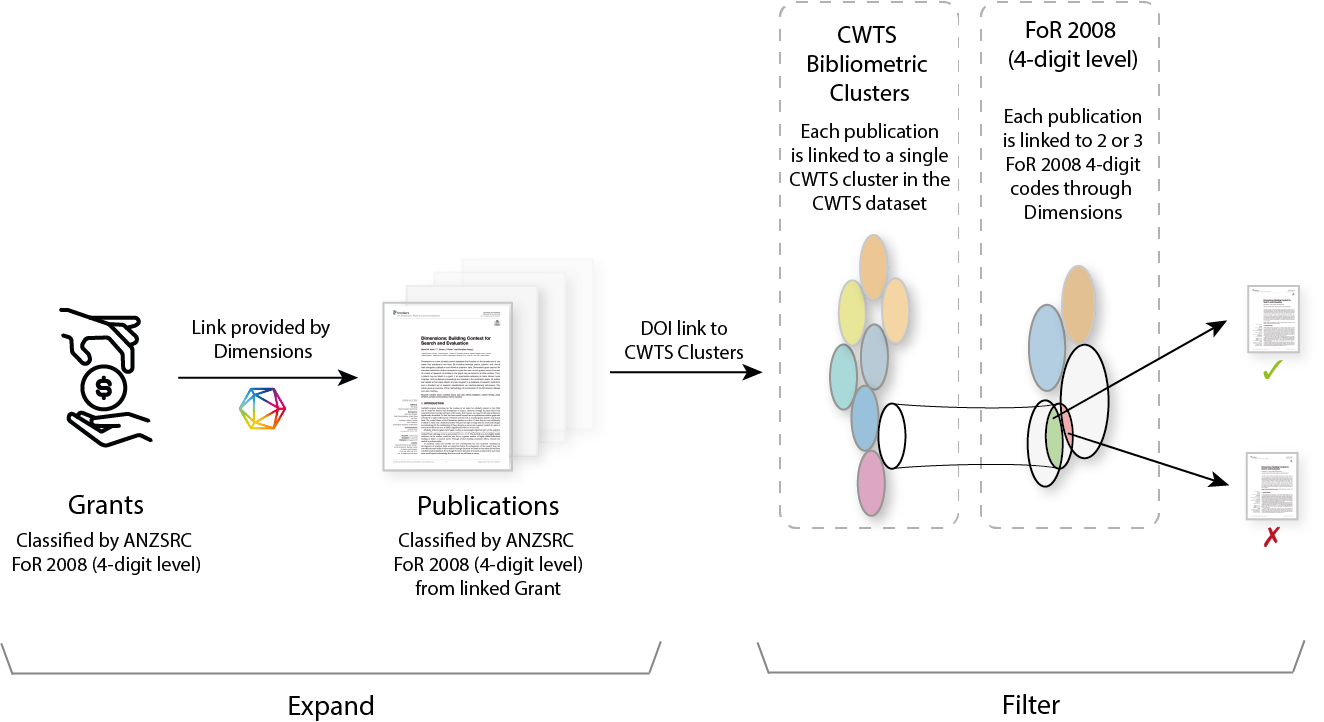}
    \caption{Illustration of strategy 1.  Grants data, which has been manually classified at the 6-digit level, is taken as a base set from which to work.  Publications that acknowledge these grants are given the 4-digit-level code of the 6-digit code associated with the grant.  These publications are then filtered to ensure that this attribution is a reasonable assumption.  The CWTS bibliometric clusters (for which their is also a DOI-to-cluster mapping for each publication) is used to ensure that the grant-affiliated code is relevant to the publication.}
    \label{fig:1}
\end{figure*}
The first strategy both broadens the dataset and moves from being grant-centric to publication-centric.  The reason for doing this is two-fold - firstly, a publication is a much richer source of metadata from which to derive a training dataset for a classification algorithm than is a grant; secondly, most grants result in more than a single publications and hence there is a natural scaling-up of the data.  In addition, publications will necessarily be more recent than the original grant that funded them and hence have more modern linguistic variants that will tend to make the dataset more representative and robust for algorithmic training.

A route to reliably annotating a publication-based training set would then give us a training set that is larger, more aligned with the types of outputs that it will most frequently be employed in classifying, and more timely.  This should allow not only the algorithm to assign codes to objects with greater confidence resulting in better coverage, but should also result in better and more reliable inter-field discrimination.

The starting point for our process is provided by the grant datasets made available by the Australian Research Council (ARC) and National Health and Medical Research Council (NHMRC) in which each grant has been assigned codes manually at a 6-digit level.  We move from this grant-centric dataset by taking each publication that is attributed to each of these grants in Dimensions and transferring the 4-digit level part of the FoR 2008 code associated with the grant to be associated with each acknowledging paper. 

We have made an implicit assumption that publications associated with a grant will share the same classifications as the grant that funded them. Even with the coarse-graining introduced by associating the 4-digit rather than the more detailed 6-digit level code, this may not be true in all cases, and we cannot argue that the dataset is large enough for erroneous affiliations to be an edge effect or statistically negligible. As a result, we devised a filtering mechanism that removes paper-classification pairings that may be incorrect.  The full process is summarised in Figure~\ref{fig:1}.

\begin{figure*}
    \centering
    \includegraphics[width=.9\linewidth]{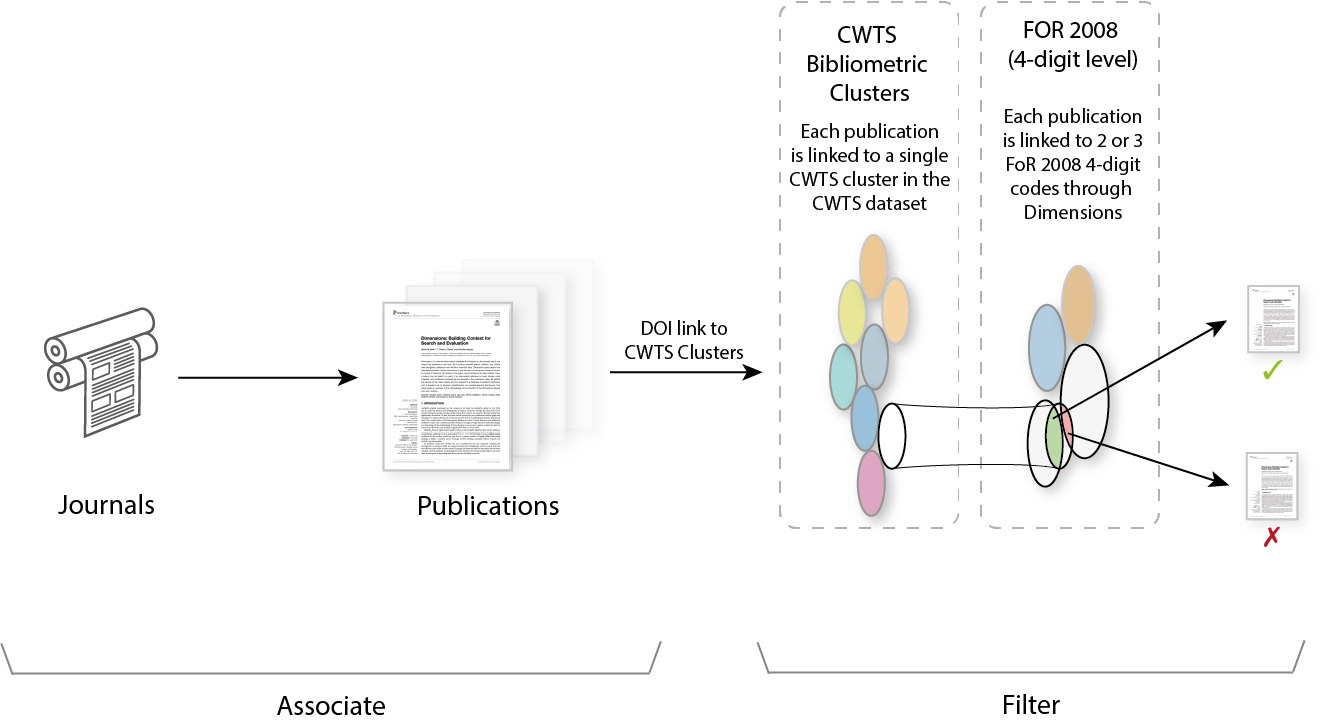}
    \caption{Illustration of strategy 2 to extend the core training data.}
    \label{fig:2}
\end{figure*}

The filtering mechanism is designed to ensure that articles gaining a code from their supporting grant are well described by that code.  The coding for the grants are assigned manually and so they should have had a good level of curation.  In this filter there are two links that we rely on - the first is a firm link from the paper in question to the CWTS bibliometric clusters \cite{waltman_unified_2010, subelj_clustering_2016} - this is provided as part of the CWTS bibliometric cluster dataset.  Thus, first we associate a selected paper, which we call Paper A, with its presumptive grant-derived FoR code, which we call Code B. We look up Paper A’s DOI to identify its CWTS bibliometric cluster, which we will call Cluster C.  We then want to check to see if Code B makes sense in combination with Cluster C. To do this, we look at the percentage of papers that Cluster C represents with respect to Code B.  The current Dimensions mapping is inaccurate, but treated statistically, it gives us a basis from which we can attempt to assess the robustness of the connection that we are trying to make between Paper A and Code B.  If more that 1\% of Code B papers are allocated to Cluster C then we allow Code B to be associated with Paper A.  This filtering mechanism allows us to improve the accuracy of the FoR ascertions for each paper in the training set.

As a technical aside, it is worth noting that the CWTS bibliometric clusters constitute a highly stable set of communities in which the connectedness of a given community is guaranteed {\cite{traag_louvain_2019}}.  For the purposes of acting as a filter in our method, this is a necessary and sufficient level of connectedness.  The overall sensitivity of our method to this step is low since we are not actively using the bibliometric clusters actively as a component of our new mapping, but rather more passively as a filter to ensure that papers are cohesive within topics.

Strategy 1 moves us from a core set of 60,000 or so grants/FoR assertions to almost 440,000 Publication/FoR assertions.  The resultant dataset is also slightly more international than the original dataset as the international collaborators of Australia-based and New Zealand-based researcher are co-authors on these papers (around 97.5\% of ARC, NHMRC, MBIE, Royal Society of New Zealand, and Health Research Council-funded papers have at least one international co-author.)

\subsection{Strategy 2: Expanding international representation}

A second strategy allows to us to extend the training dataset further.  For clarity, while Strategy 2 follows a similar structure to Strategy 1, this strategy departs completely from the previous one in that it does not relate to the original grant dataset on which Strategy 1 was based.  Rather, this strategy provides a distinct additional set of training data.

Figure~\ref{fig:2} is a schematic representation of Strategy 2.  The fundamental observation that underlies this strategy is that many journals contain a general name for their field - for example, “Journal of Physics A: Mathematical and Theoretical”, some are even more precisely aligned with FoR codes.  While this may seem tenuous, it is rather a rich seam to exploit.  For example, FoR 1108 is ``Medical Microbiology'' for which there is not only the Journal of Medical Microbiology (published by the Microbiology Society), but 18 further journals are listed in Dimensions that contain the term in their title, ranging from the expected geographical variants such as the International Journal of Medical Microbiology (published by Elsevier) and the Egyptian Journal of Medical Microbiology (published by the Egyptian Society for Medical Microbiology) to journals that work across fields such as the Journal of Medical Microbiology and Infectious Diseases (published by the Pasteur Institute of Iran).  In cases where articles appear in these types of journal we have a strong indication of a likely FoR code that should be applied to the article.

There are many cases where such journals publish articles that are either not centrally in the field or are tangential to the core field (for example, the publication of bibliometric or social commentary on the development of the field).  In these cases, we can employ the same filtering mechanism that we employed in Strategy 1 to ensure that the dataset is well aligned with the core of the area that we wish to model.

\subsection{Strategy 3: Community contributions}

To supplement strategies 1 and 2, and pushing back a little toward a geographic bias, we invited universities in the Australian sector to contribute any data that they felt comfortable with donating to our mapping project.  Two universities who had already undertaken significant manual mapping exercises were kind enough to supply their data, coded at 6-digit level, to help our effort.

Strategy 3 provided just over 130,000 further publication/FoR assertions.

\subsection{Strategy 4: Mapping}

A further strategy leverages the correspondence table between FoR-2008 and FoR-2022 codes provided by the ARC {\cite{arc_australian_2020}}. Using the datasets for which the 2008 6-digit-level coding was available, we identified CWTS clusters where, within a given cluster, publications from a 2008 4-digit-FoR code map to a single FoR-2020 code.  This approach goes beyond making a simple one to one mapping between schemes, by allowing a cluster to effectively split mappings.  An FoR-2008 code might map to a different FoR-2020 code depending on the cluster with which it is associated. Using these rules in combination with the existing FoR-2008 codes already available within Dimensions, a further 4.7million publications can be mapped.

A much smaller range of codes did not fit into either of these first two classifications, which then meant examining which journals were likely to map to specific 2020 codes, and then doing filter tests such as those described in Strategies 1 and 2.

Finally, for 81 6-digit-FoR codes and for 42 4-digit-FoR codes, which defied any of the previous strategies, we defined specific search terms based on title and abstract searches, to build reference corpora for these fields.  This a time-consuming task and requires expert input, but is a methodology that we have employed before in constructing SDG classifications \cite{wastl_contextualizing_2020}.  This strategy yielded an additional 330,000 publication/FoR connections.

We then retained coding experts in the Australian sector to review key problem areas of the classification, to manually review and curate the mapping.  Simultaneously, we made our initial categorisation outputs available in an online tool for the sector to feedback on. As a result of this community interaction, we also received a dataset from a third Australian university, who kindly analysed and tested our results against their own internal dataset, which lead to further improvements in the final mapping for the training dataset.

\subsection{Summary of machine-learning approach}

The classification system in Dimensions is built upon supervised machine learning, whereby manually coded examples are used to train a model that then serves to automatically predict uncoded data (such as publications, grants, clinical trials, patents and reports). The classifier follows the classic pipeline of TF-IDF {\cite{Salton88}} scoring over word n-grams that are used to produce a linear model with the support vector machines  (SVM) {\cite{Cortes95}}. With this setup, the classifier is capable of returning a ranked and scored list of categories for each input document. It is a fairly simple and fast, yet effective setup commonly used in text classification ({\cite{Joachims98}} and {\cite{Yang99}}).

The model is then evaluated for consistency. This is performed via  3-fold cross-validation where the classifier is trained and validated using two-thirds of the set and tested on the remaining third. This process is repeated 3 times, each time with different samples in the training and test sets to avoid selection bias.

\section{Results}
The classification of objects across all object types in Dimensions with the 2020 FoR codes was included in Dimensions on Google BigQuery in the 2022/08 release \cite{dimensions_release_2022}.  This new classification was the result of the methodology described above together with a process by which the training dataset was implemented into Dimensions' machine learning pipeline.  An FoR 2020 classifier API endpoint already exists. A future release of Dimensions will move to the FoR 2020 codes in the web interface.
\label{sec:3}
\begin{figure}
    \centering
    \includegraphics[width=1\linewidth]{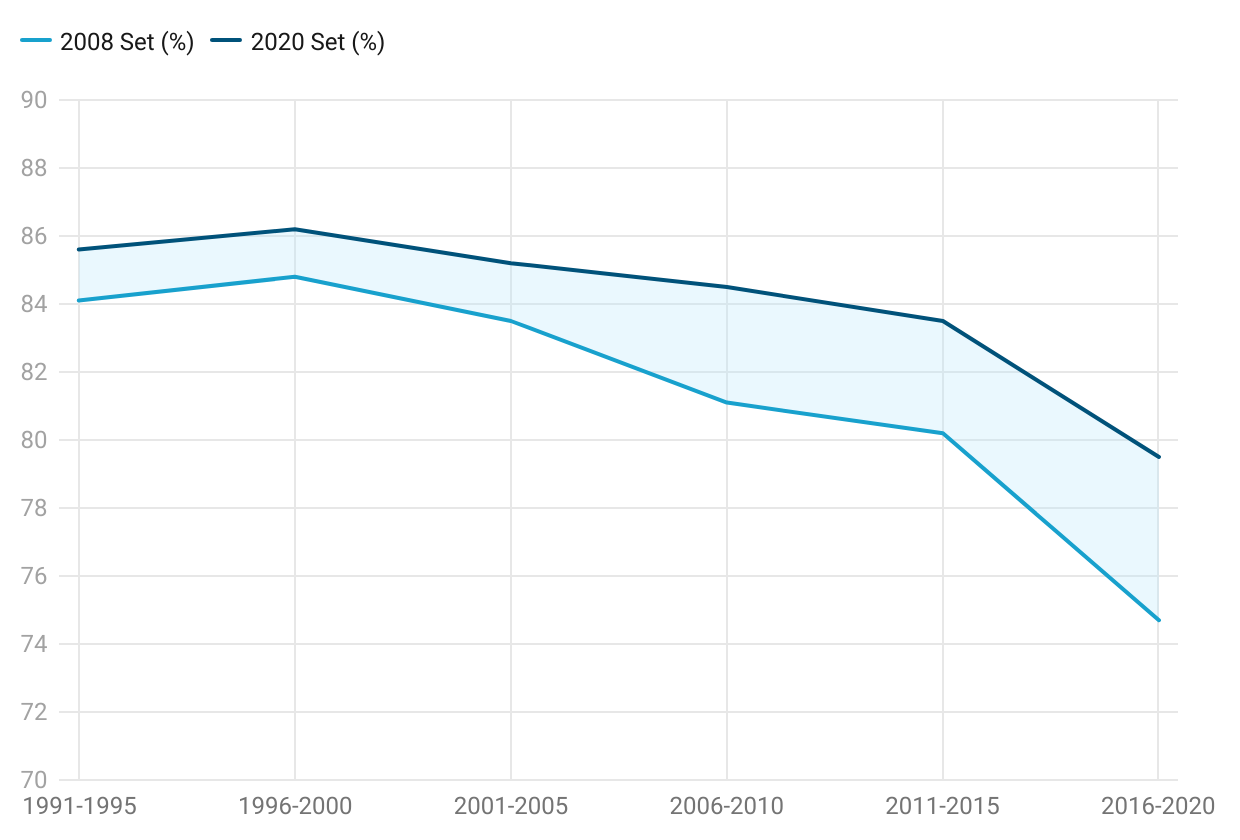}
    \caption{Coverage of FoR Codes (2008 light blue/old training set, 2020 dark blue/new training set) by 5-year grouping from 1991 to 2020.}
    \label{fig:3}
\end{figure}
\begin{table*}[]
\begin{tabular}{@{}rrrrrr@{}}
\toprule
                                & Total in Dimensions & Coverage in FoR 2008 & FoR 2008 (\%) & Coverage in FoR 2020 & FoR 2020 (\%) \\ \midrule
Clinical Trials                 & 720,554             & 617,213              & 85.7          & 716,918              & 99.5          \\
Grants                          & 6,265,754           & 4,069,292            & 64.9          & 4,248,411            & 67.8          \\
Patents                         & 148,033,299         & 97,696,637           & 66.0              & 101,378,710          & 68.5          \\
Policy Documents                & 877,912             & 781,817              & 89.1          & 807,258              & 92.0          \\
Publications (All Types)        & 129,705,024         & 94,195,766           & 72.6          & 99,362,025           & 76.6          \\
Publications (Journal Articles) & 104,875,479         & 81,248,351           & 77.5          & 84,665,668           & 80.7          \\ \bottomrule
\end{tabular}
\caption{Coverage of FoR Codes in Dimensions for different data types.}
\label{tab:1}
\end{table*}

Of the 104.8m research articles in Dimensions at the time of writing, not all articles receive an FoR code, however, the new approach detailed above significantly improves the coverage. The previous version of the FoR categorisation machine learning approach was able to classify 81.3m papers with at least one FoR code (around 77.5\% of Dimensions' publication holdings), whereas the new machine learning approach is able to increase coverage to 84.7m papers with at least one assigned FoR code (80.7\% of Dimensions' publications holdings). Figure~\ref{fig:3} shows the development of coverage from year to year between the FoR 2008 codes (old machine learning training set) and the FoR 2020 codes (new machine learning training set developed as described above). Not that the shape of the coverage percentage curve is similar in both cases - this is emblematic of the characteristics of the underlying dataset.  However, as might be expected with the approach described above, the new machine-learning model shows itself increasingly superior on more recent articles as it is better able to account for linguistic drift.  It is also noteworthy that the upper curve (dark blue, FoR 2020/new machine learning set) is not only consistently above the light blue curve but also that the level of variation in the curve is suppressed, indicating an increased reliability in the mapping.  

A similar analysis can be performed for grants information in the Dimensions database and shows that of 6.2m grant records, 4.1m (or 64.5\%) receive an FoR code under the old approach compared with 4.25m (or 67.8\%) under the new approach.  For patents, we also see parallel results with there being 66\% coverage of the 148m records using the old approach and an improvement to 68.5\% with the new approach.  Table~\ref{tab:1} summarises the results across the Dimensions database.  In each cases, we see an improvement in coverage with the new algorithm.

\begin{figure}
    \centering
    \includegraphics[width=1\linewidth]{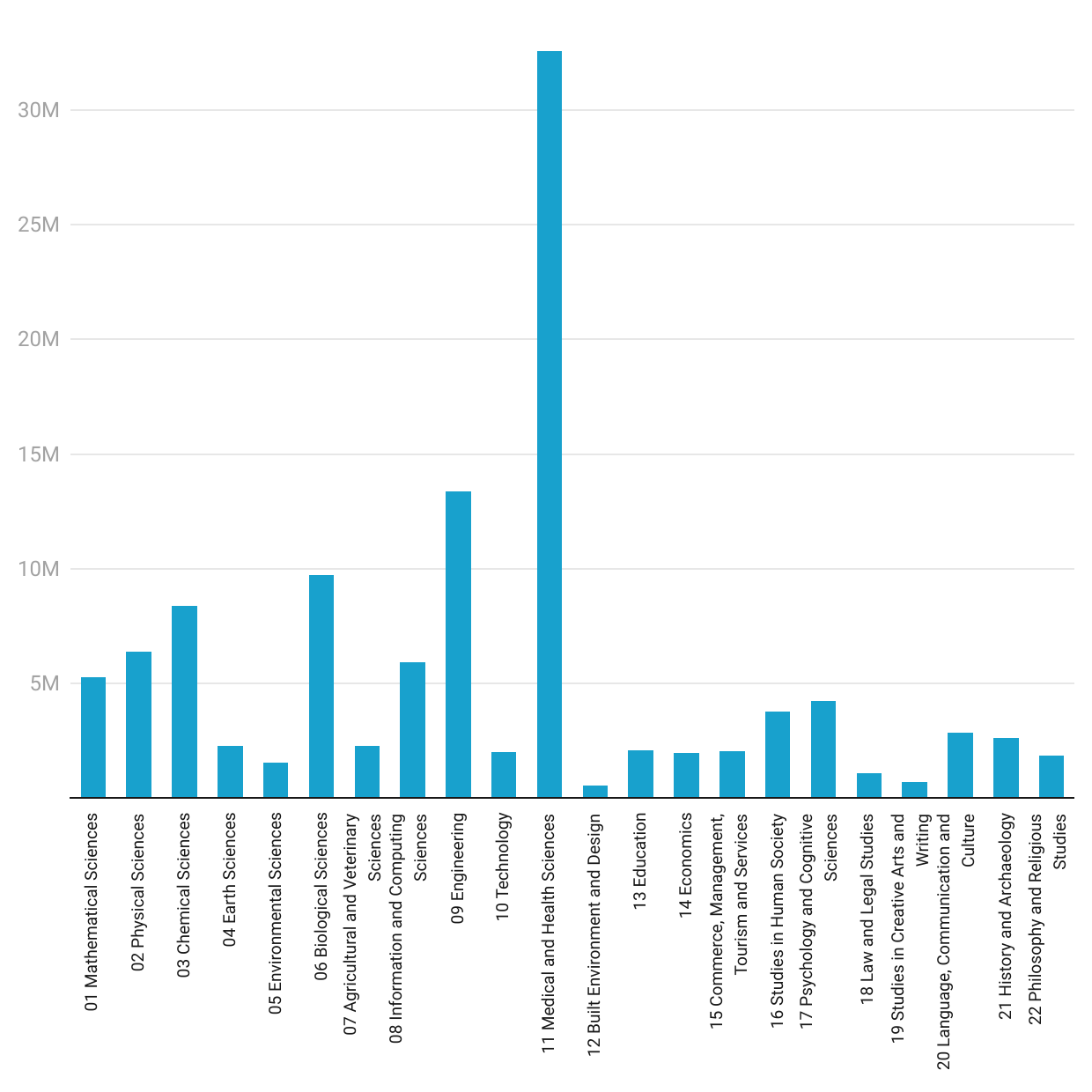}
    \caption{Population of each 2008 FoR code by number of outputs in Dimensions (articles, proceedings, preprints, chapters and monographs) for all time.}
    \label{fig:4}
\end{figure}

\begin{figure}
    \centering
    \includegraphics[width=1\linewidth]{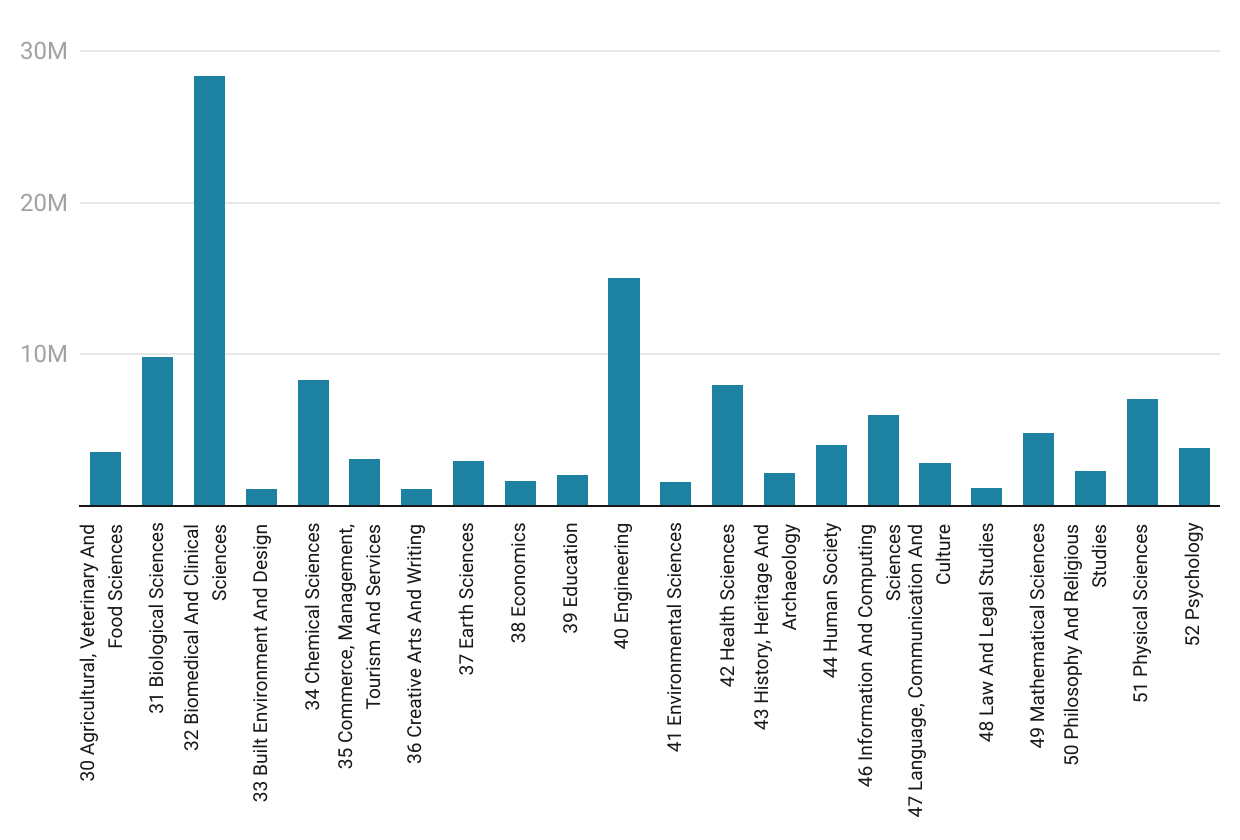}
    \caption{Population of each 2020 FoR code by number of outputs in Dimensions (articles, proceedings, preprints, chapters and monographs) for all time.}
    \label{fig:5}
\end{figure}
By comparing figures~\ref{fig:4} and \ref{fig:5} we can observe the difference between the distributions of classification affiliations between disciplines using the two different coding schemes.  It is notable that the new subject groupings lead to a distribution that is slightly more homogeneous while the overall shape remains quite stable.  The most significant shift is in the codes associated with medicine-related codes where 28.7\% of affiliated outputs (articles, proceedings, preprints, chapters and monographs) having a link to FoR 2008 code \textit{11 Medical and Health Sciences} to just 23.5\% of affiliated outputs having a link to FoR 2020 code \textit{32 Biomedical And Clinical Sciences} with \textit{42 Health Sciences} being broken out into a separate code accounting for 6.6\% of affiliations in the FoR 2020 Codes.

\begin{figure*}
    \centering
    \includegraphics[width=1\linewidth]{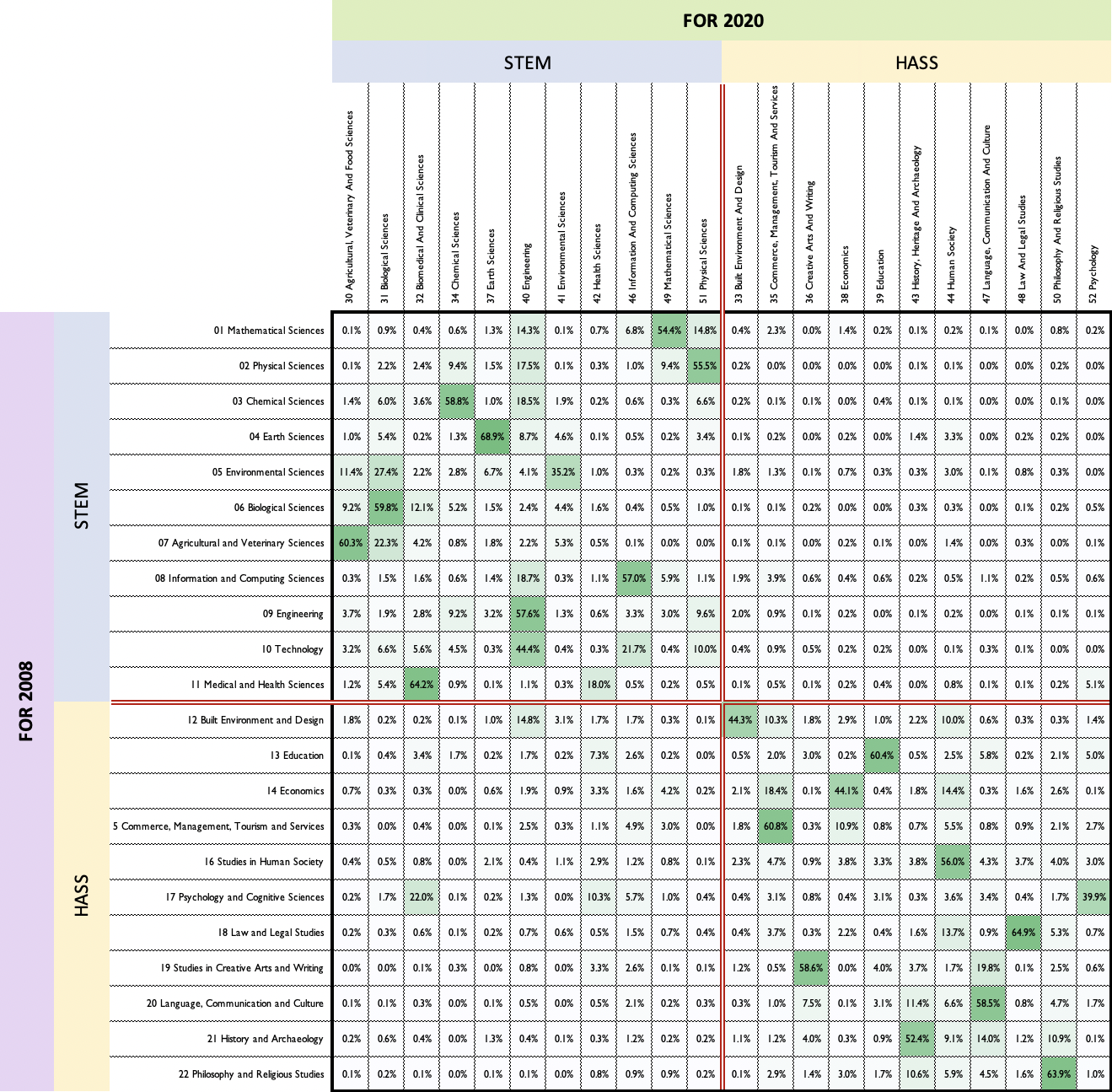}
    \caption{Mapping from FoR 2008 codes to FoR 2020 codes in Dimensions.  \change{Each row shows the percentage of each of the FoR 2008 code that have been remapped to each of the new FoR 2020 codes.}{The FoR 2008 codes correspond to rows and the FoR 2020 code to columns. Each cell shows the percentage overlap between old codes and new codes.  The sum of each row must be 100\%, the sum of each column must be 100\%.} The stronger green colouring corresponds to a higher percentage overlap.  The table has been grouped by STEM (Science, Technology, Engineering and Medicine) and HASS (Humanities, Arts and Social Sciences).  This mapping is, of course, fuzzy as several categories that we have classified as STEM might be argued to be HASS and vice versa.  However, it is noteworthy that the upper-right and the lower-squares (delineated by double red lines) have extremely low percentages, this indicates low cross mapping from STEM to HASS, as might be hoped for.}
    \label{fig:6}
\end{figure*}

Figure~\ref{fig:6} summarises the percentages of articles mapped from a given 4-digit-FoR-2008 code across all 4-digit-FoR-2020 codes.  Note that even in simple cases, this is not a one-to-one mapping with 100\% of articles being mapped from one coding to another.  There are two major drivers of this: Firstly, the fact that the mapping from the 2008 FoR codes is not a simple mapping exercise where one code took over from another - there is a clear (and needed) update of research areas that better represents the current state of the art and a consequent redistribution of research outputs amongst new codes; secondly, there is an improvement in our the core machine-learning training set as a result of the process above.  Hence, we assert that the discrimination of the algorithm has been improved to better classify articles more accurately.

The confluence of these effects can be seen in almost every mapping, for example, \textit{01 Mathematical Sciences} is redistributed with 54.4\% being mapped to \textit{49 Mathematical Science}s and around 14\% being reclassified each to of \textit{40 Engineering} and \textit{51 Physical Sciences} respectively, with reciprocal realignments mapping 9.4\% of\textit{ 02 Physical Sciences} and 3\% of \textit{09 Engineering} being attributed back to \textit{49 Mathematical Sciences}.  This type of movement will be substantially due to improvements in the algorithm with some natural movement in fields over a decade.  However, \textit{17 Psychological and Cognitive Sciences} was clearly split between \textit{32 Behavioural and Clinical Sciences} and \textit{52 Psychology}, with some spillover into \textit{42 Health Sciences}, as can clearly be seen in the data. FoR \textit{10 Technology} also has no direct mapping in the FoR 2020 codes and has been substantially shared between \textit{40 Engineering} and \textit{46 Information and Computing Science}.

\section{Discussion}
\label{sec:4}
\subsection{Limitations and challenges}
There are a number of obvious limitations with this approach that  fall into two broad categories - i) issues with data limitations and ii) issues with machine learning.

In the first category, there is a clear methodological limitation in that we have employed a kind of circular logic in our model where we are using our existing machine-learned mappings in order to filter out noise in the new model. This naturally means that there is a heavier influence on the new model by the previous model than would be ideal.  However, care was taken to set the filter at a level to remove real outliers, rather than focus on the core of an area.  Broadly, we deem this approach to have been quite effective.

Another challenge was the level of data that we could obtain to model the indigenous categories.  In fact, so sparse was the data, that we could not make a large enough dataset to create a machine-learning training set for these areas.  The challenge here is that the new indigenous categories are quite fine-grained while simultaneously being specific to Australia and New Zealand. (While indigenous research does take place in many other places in the world, its nomenclature and references tend to be centred around different geographies and hence it is broadly unhelpful in creating a generalised mapping for indigenous topics.) For this reason, indigenous research is not included in our machine learning model.

A further data challenge rests in some of the sources of the data and the way in which humans have categorised the data.  Data, both from institutional partners and through research council sources, is tagged from an intrinsically ``top down'' perspective in the sense that the categorisation system has been provided for researchers and university administrators to match to and that matching comes with certain incentives and biases around funding optimisation.  We do not point this out as necessarily a bad thing, but rather that it is an aspect of the schema and the categorisation that should be transparent to the user as it will have a influence on whether the categorisation is well suited to certain use cases.  While the Strategy 4 is more of a ``ground up'' strategy, it is nonetheless, only a component of the overall mix, which is dominated by top-down-influenced data.

Without context any categorisation system is dangerous since the ``lens'' mentioned in our introduction is not well understood.  Since the system discussed in this paper will specifically be used in Dimensions it is noteworthy that the majority of users will encounter the categorisation system without the deeper context provided by this paper.  At one level, given the nature of how this categorisation has been created, there are clearly some challenges with its use at either individual levels needed for evaluative settings, but at another level, there is a value to this system at a mass reporting level.  Even at the individual publication level, we would argue that there is a value in producing a homogenised, non-subjective, basis for discussion and disagreement and that, perhaps, the Dimensions categorisation is a route to opening up discussions that might otherwise not be possible. In any case, we believe that sharing the current methodology is critical to help inform users who wish to make use of this categorisation scheme in the context of Dimensions.

In the second category, concerning issues around machine learning, our early models based on our newly extended and mapped data clearly learned from facets of the data that we had not anticipated. In particular, each publication in our training set only had a single FoR code associated with it, which meant that each publication classified by early versions of the new model only supplied a single classification.  Another issue was that the model appeared to have learned that each field should be the same size, as we had tried to create a balanced dataset with equitable representation across FoR codes.  This led to articles being mis-classified as there were ``too many'' existing articles classified into particular codes.  This meant that some work was needed to statistically shape the learning dataset so that it was more representative of production levels in different areas.

\subsection{Outlook and implementation}
As we reflected in the opening section to this paper, the scope of a classification system is vast.  Not only are there significant challenges in designing and defining a classification system to meet the needs of both its divisors and users, there are also huge structural and infrastructural challenges in adopting and implementing such a system and then maintaining and updating it.  In this paper we have touched on a few of those issues in a detailed manner to illustrate the process of updating Dimensions - a search, discovery and analysis tool.  Across the sector, however, there are many more systems that depend on classification systems: research information management solutions, grant application systems, data repositories, and so on.  While the present use case is likely to be among the more complex ones, it is certain that more work needs to be done on how research classification can evolve in a sustainable and consistent manner.

We note that in previous years the ARC has made available a Journal List for the Excellence in Research for Australia (ERA) evaluation exercise where each journal received up to three 4-digit-FoR-code classifications.  It is, of course, possible to derive such a listing statistically from the classification work that we have done here by awarding journals the 3 FoR Codes that most appear in recent publications published in a given journal with some kind of time cutoff to ensure good agreement with the most recent research.

Designing a system that is capable of being updated in a more continuous and responsive way could be  significant from a systems perspective, that can pose challenges and is  in tension with the needs of evaluators. As such, we believe that more thought needs to be put into classification systems and how they are practically deployed, in order to serve the changing information needs of the sector. 

\section*{Acknowledgements}
The authors wish to thank Mario Diwersy for his support in developing this approach and their translation of the training data that came out of this work into a functioning  categorisation in Dimensions.  Additionally, the invaluable input and assistance of Jo Dalvean and Merren Cliff as local experts in the FoR categorisation approaches used in an evaluative context in the Australian sector.  We are also indebted to several institutions who provided feedback on early versions of the classification scheme.  Thanks go to Anne Harvey and Danu Poyner for their ongoing efforts to engage with Australasian stakeholders to gather feedback, without which this work would not have been possible.  Finally, we thank the team at CWTS for making their bibliometric clusters available for this work.

\section*{Data availability statement}
The data for Figures~\ref{fig:3}-\ref{fig:6} are available from Figshare at \href{https://doi.org/10.6084/m9.figshare.20691838}{https://doi.org/10.6084/m9.figshare.20691838}. A version of Dimensions is made freely available for personal use that allows the validation and checking of certain aspects of this paper. However, at the time of writing, to investigate further an academic licence (see {\cite{herzog_dimensions_2020}}) or subscription would be required.

\section*{Conflict of interests}
DWH, LH and SJP are employees of Digital Science, the owner and operator of Dimensions. It is important to note that, from the standpoint of reproducibility, Dimensions is a commercial product and hence it has not be possible to provide complete transparency for reviews to gain an insight into the full methodology behind certain aspects of this paper.

\section*{Author contribution statement}
Conceptualization: SJP, DWH; Data curation: SJP; Formal Analysis: DWH; Investigation: SJP; Methodology: SJP, LH; Project administration: SJP; Software: LH; Validation: LH; Writing - original draft: DWH; Writing - review and editing: DWH, SJP, LH.

\bibliography{FORs}

\end{document}